\documentclass[preprint,preprintnumbers, prd, floatfix, superscriptaddress,nofootinbib] {revtex4-1}
\usepackage{graphicx}
\usepackage{epstopdf}
\usepackage{epsfig}
\usepackage{subfigure}
\usepackage{dcolumn}
\usepackage{bm}
\usepackage[usenames ,dvipsnames]{xcolor}
\usepackage{slashed}
\usepackage{graphicx,color}
\begin{document}
\title{Cabibbo-favored $\Lambda^+_c\to\Lambda a_{0}(980)^+$ decay\\
in the final state interaction}

\author{Yao Yu}
\email{yuyao@cqupt.edu.cn}
\affiliation{Chongqing University of Posts \& Telecommunications, Chongqing 400065, China}

\author{Yu-Kuo Hsiao}
\email{yukuohsiao@gmail.com (Corresponding author)}
\affiliation{School of Physics and Information Engineering, Shanxi Normal University, Linfen 041004, China}

\begin{abstract}
The anti-triplet charmed baryon decays with the light scalar mesons are rarely measured,
whereas the recent observation of the Cabibbo-favored $\Lambda_c^+\to \Lambda\eta\pi^+$ decay
hints a possible $\Lambda_c^+\to\Lambda a_0(980)^+,a_0(980)^+\to \eta\pi^+$ process.
We hence study the $\Lambda_c^+\to\Lambda a_0(980)^+$ decay. Particularly,
it is found that  the final state interaction can give a significant contribution,
where $\Sigma^{+}(1385)$ and $\eta$ in $\Lambda_c^+\to \Sigma^{+}(1385)\eta$
by exchanging a charged pion are transformed as $\Lambda$ and $a_0(980)^+$, respectively.
Accordingly, we predict
${\cal B}(\Lambda_c^+\to\Lambda a_0(980)^+)=(1.7^{+2.8}_{-1.0}\pm 0.3)\times 10^{-3}$,
accessible to the BESIII, BELLEII and LHCb experiments.

\end{abstract}

\maketitle
\date{\today}
\section{Introduction}
For the light scalar meson ($S_0$),
such as $(a_0,f_0)\equiv (a_0(980),f_0(980))$~\cite{pdg},
it is controversial that one can regard
its structure as a $q\bar q$, impact tetraquark,
or meson-meson bound state~\cite{review}. For clarification,
more measurements with the $S_0$ states are needed.
Most decay channels with $S_0$ have been observed in the $B$ and $D$ meson decays.
On the other hand,
it is also possible that $S_0$ can be produced in the charmed baryon decays,
such as ${\bf B}_c\to {\bf B}S_0$,
where ${\bf B}_{(c)}$ denotes the octet (anti-triplet charmed) baryon.
Nonetheless, except for
${\cal B}(\Lambda_c^+\to p f_0)=(3.5\pm 2.3)\times 10^{-3}$ measured in 1990~\cite{Barlag:1990yv,pdg},
there is no ${\bf B}_c\to {\bf B}S_0$ to be newly observed.

To seek the new decay channels of ${\bf B}_c\to {\bf B}S_0$,
one has considered the resonant decay
$\Lambda_c^+\to pa_0^0,a_0^0\to K^+K^-(\eta\pi^0)$~\cite{Wang:2020pem,Li:2020fqp},
where the estimation gives ${\cal B}(\Lambda_c^+\to pa_0^0)\sim 10^{-5}-10^{-4}$.
This might still cause a difficult measurement. Recently,
BELLE has reported the observation of
the Cabibbo-favored three-body $\Lambda_c^+\to \Lambda\eta\pi^+$ decay~\cite{Lee:2020xoz},
which can help to explore the $a_0^+$ state~\cite{Xie:2016evi}.
Explicitly, the total branching fraction
is given by~\cite{Lee:2020xoz}
\begin{eqnarray}\label{data1}
{\cal B}(\Lambda_c^+\to \Lambda\eta\pi^+)=
(18.4\pm 0.2\pm 0.9\pm 0.9)\times 10^{-3}\,,
\end{eqnarray}
which is found to receive the resonant contributions from
\begin{eqnarray}\label{data2}
{\cal B}(\Lambda_c^+\to \Lambda^*\pi^+,\Lambda^*\to\Lambda\eta)
&=&(3.5\pm 0.5)\times 10^{-3}\,,\nonumber\\
{\cal B}(\Lambda_c^+\to \Sigma^{*+}\eta,\Sigma^{*+}\to\Lambda\pi^+)
&=&(10.5\pm 1.2)\times 10^{-3}\,,
\end{eqnarray}
with $\Lambda^*\equiv\Lambda(1670)$ and $\Sigma^*\equiv\Sigma(1385)$.
Most interestingly, the Dalitz plot of Fig.~5 in~Ref.~\cite{Lee:2020xoz} presents
an ambiguous band,
suggesting a possible $\Lambda_c^+\to\Lambda a_0^+,a_0^+\to \eta\pi^+$ process.

\begin{figure}[t!]
\includegraphics[width=3.0in]{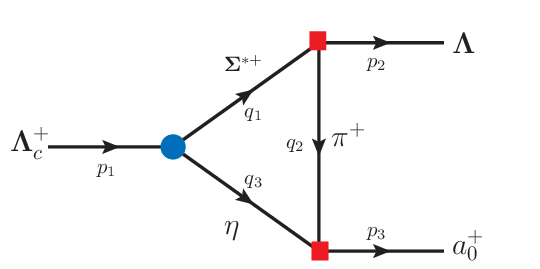}
\caption{The triangle rescattering process for
the two-body $\Lambda_{c}^+\to\Lambda a^{+}_{0}$ decay.}\label{triangle}
\end{figure}

With the possible signal,
the $\Lambda_c^+\to \Lambda a_0^+$ decay is worth a careful investigation.
Theoretically, its branching fraction has been estimated as
small as $1.9\times 10^{-4}$,
which is based on the factorization and pole model
that deal with the factorizable and non-factorizable effects~\cite{Sharma:2009zze}.
By contrast,
the rescattering effect for $\Lambda_c^+\to\Lambda a_0^+$ has not been investigated yet.
See Fig.~\ref{triangle},
$\Sigma^{*+}\eta$ from $\Lambda_c^+\to \Sigma^{*+}\eta$
can be transformed as $\Lambda a_0^+$,
which is by exchanging a charged pion.
Since ${\cal B}(\Lambda_c^+\to \Sigma^{*+}\eta)$ is
at the level of $10^{-2}$~\cite{Lee:2020xoz,pdg},
and the strong decays of $\Sigma^{*+}\to \Lambda\pi^+$ and $a_0^+\to \eta\pi^+$
are not small~\cite{pdg}, we expect that the triangle rescattering process
can significantly contribute to $\Lambda_c^+\to \Lambda a_0^+$,
instead of the (non-)factorizable effects~\cite{Zhao:2018mov,Hsiao:2020iwc,Pan:2020qqo,Kohara:1991ug}.
In this report, we will calculate the triangle rescattering in Fig.~\ref{triangle},
and predict the branching fraction of $\Lambda_c^+\to\Lambda a_0^+$,
to be compared to the future measurements.
We will also test other contributions with
the triangle singularity~\cite{Liu:2019dqc,TS,Guo:2019twa}.
\begin{figure}[t!]
\includegraphics[width=2.8in]{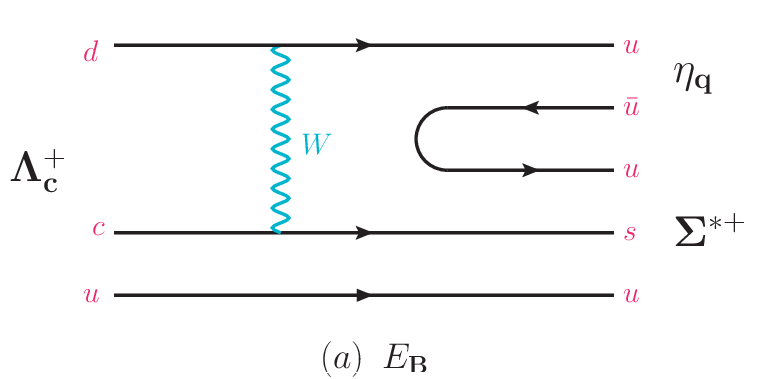}
\includegraphics[width=2.8in]{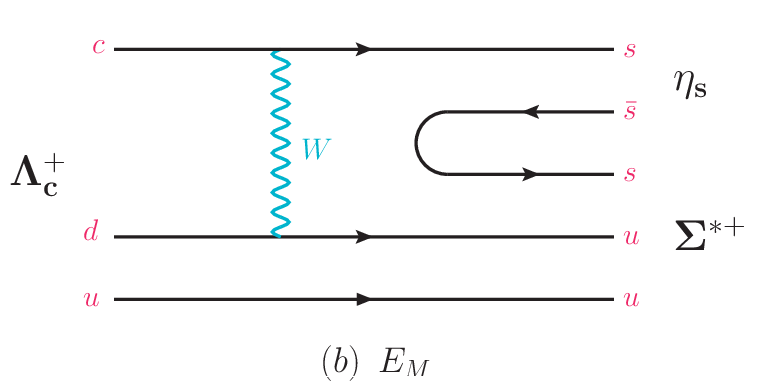}
\caption{The $\Lambda_c^+\to\Sigma^{*+}\eta$ weak decay
proceeds through the topological diagrams in (a,b), which are
are parameterized as $(E_{\bf B},E_{M})$, respectively.}\label{EE}
\end{figure}
\section{Formalism}
As the final state interaction, the triangle rescattering process has been
well applied to $D_s^+\to \pi^{+(0)} a_0^{0(+)}$ and
$\Lambda_c^+\to \Sigma^{+(0)}\pi^{0(+)}$~\cite{Hsiao:2019ait,Ke:2020uks}.
Likewise, we propose the rescattering
$\Lambda^{+}_{c}\to\Sigma^{*+}\eta\to\Lambda a_{0}^{+}$ decay
as depicted in Fig.~\ref{triangle},
which involves the weak decay of $\Lambda^{+}_{c}\to \Sigma^{*+}\eta$,
and the strong decays of
$\Sigma^{*+}\to \Lambda\pi^{+}$ and $a_0^+\to\eta\pi^+$.
In particular,
the weak $\Lambda^{+}_{c}\to \Sigma^{*+}\eta$ decay
is found to include the two $W$-boson exchange processes in Fig.~\ref{EE}~\cite{Kohara:1991ug}.
Specifically,
Fig.~\ref{EE}a(b) presents the topology for
the baryon (meson) to receive the $s$ quark in the $c$ to $s$ transition,
which can be parameterized as $E_{{\bf B}(M)}$ in the topological diagram scheme~\cite{Hsiao:2020iwc}.
To proceed,
the corresponding amplitudes can be written
as~\cite{Hsiao:2020gtc,Hsiao:2019ait,Xie:2017xwx}
\begin{eqnarray}\label{3M}
{\cal M}_1&\equiv& {\cal M}(\Lambda^{+}_{c}\to \Sigma^{*+}\eta)=
\frac{G_F}{\sqrt 2} V^*_{cs}V_{ud}\epsilon_\mu
\bar{u}^{\mu}_{\Sigma^*}(a_E-b_E\gamma_{5})u_{\Lambda^{+}_{c}}\,,\nonumber\\
{\cal M}_2&\equiv& {\cal M}(\Sigma^{*+}\to \Lambda\pi^{+})=
g_{\Sigma^* \Lambda\pi}\epsilon_\mu \bar u_\Lambda u^{\mu}_{\Sigma^*}\,,\nonumber\\
{\cal M}_3&\equiv& {\cal M}(a_0^+\to\eta\pi^+)=g_{a_{0}\eta\pi}\,,
\end{eqnarray}
with $\epsilon_\mu\equiv q_\mu/\sqrt{q^2}$~\cite{Gutsche:2018utw},
where $q_\mu$ is the four-momentum of the meson,
$u^{(\mu)}$ the spinor of the spin-1/2 (3/2) baryon, and
$(g_{\Sigma^* \Lambda\pi},g_{a_0\eta\pi})$ the strong coupling constants.
In the quark-diagram scheme, it is obtained that
$a_E(b_E)=\sqrt {1/6}(-E_{\bf B}c\phi+\sqrt 2 E_M s\phi)$
with $(s\phi,c\phi)\equiv (\sin\phi,\cos\phi)$~\cite{Hsiao:2020iwc}, where
the angle $\phi=(39.3\pm1.0)^\circ$ is from the $\eta-\eta^\prime$ mixing matrix,
given by~\cite{FKS}
\begin{eqnarray}\label{eta_mixing}
\left(\begin{array}{c} \eta \\ \eta^\prime \end{array}\right)
=
\left(\begin{array}{cc} \cos\phi & -\sin\phi \\ \sin\phi & \cos\phi \end{array}\right)
\left(\begin{array}{c} \eta_q \\ \eta_s \end{array}\right)\,,
\end{eqnarray}
with $\eta_q=\sqrt{1/2}(u\bar u+d\bar d)$ and $\eta_s=s\bar s$.

The rescattering amplitude of
$\Lambda_{c}^{+}(p_1)\to\Sigma^{*+}(q_1)\eta(q_3) \to\Lambda(p_2) a_{0}^{+}(p_3)$
with the exchange of $\pi^+(q_2)$ is in accordance with the momentum flows in Fig.~\ref{triangle}.
Using the Cutkosky rule,
we present the rescattering amplitude as~\cite{Ke:2020uks,Li:1996yn,Cheng:2004ru}
\begin{eqnarray}\label{Aab2}
&&{\cal M}(\Lambda_{c}^+\to \Lambda a_{0}^{+})\nonumber\\
&&=\frac{1}{2}\int
\frac{d^3\vec{q}_{1}}{(2\pi)^{3}2E_{1}}
\frac{d^3\vec{q}_{3}}{(2\pi)^{3}2E_{3}}
(2\pi)^{4}\delta^4(p_{1}-q_1-q_3)
{\cal M}_1 {\cal M}_2 {\cal M}_3 \frac{ F^{2}(q_{2}^2)}{q^{2}_{2}-m^{2}_{\pi}}\nonumber\\
&&=\frac{1}{2}\int
\frac{d^3\vec{q}_{1}}{(2\pi)^{3}2E_{1}}
\frac{d^3\vec{q}_{3}}{(2\pi)^{3}2E_{3}}
(2\pi)^{4}\delta(p_{1}-q_1-q_3)
\frac{g_{\Lambda_{c}\Sigma^{*}\eta}g_{\Sigma^{*}\Lambda\pi}g_{a_{0}\eta\pi}}{m_\eta m_\pi}
\nonumber\\
&&\times
\bar{u}_\Lambda q^{\mu}_{2}u_{\mu\Sigma^*}
\bar{u}_{\nu \Sigma^*}q^{\nu}_{3}(1-\gamma_{5})u_{\Lambda_c}
\frac{ F^{2}(q_{2}^2)}{q^{2}_{2}-m^{2}_{\pi}}\nonumber\\
&&=\int \frac{|\vec{q}_{1}|d\Omega}{32\pi^{2}}
\frac{g_{\Lambda_{c}\Sigma^{*}\eta}g_{\Sigma^{*}\Lambda\pi}g_{a_{0}\eta\pi}}{m_{\Lambda_c} m_\eta m_\pi}
q^{\mu}_{2}q^{\nu}_{3}
\bar{u}_\Lambda P_{\mu\nu}(1-\gamma_{5})u_{\Lambda_c}
\frac{ F^{2}(q_{2}^2)}{q^{2}_{2}-m^{2}_{\pi}}\,,
\end{eqnarray}
with $g_{\Lambda_c\Sigma^*\eta}\equiv (G_F/\sqrt 2)V^*_{cs}V_{ud}a_E$
and $E_{1(3)}$ denoting the energy of $\Sigma^{*+}(\eta)$.
For $d\Omega\equiv 2\pi \,d\cos\theta$,
$\theta$ is the angle between $\vec{q}_1$ and $\vec{p}_2$,
where $\vec{p}_{2}$ is fixed in the z direction.
Besides, we present $|\vec{q}_1|$ as
\begin{eqnarray}
|\vec{q}_1|=
\frac{1}{2m_{\Lambda_{c}}}(m^{4}_{\Lambda_{c}}
+m^{4}_{\Sigma^*}+m^{4}_{\eta}-2m^{2}_{\Lambda_{c}}m^{2}_{\Sigma^*}
-2m^{2}_{\Lambda_{c}}m^{2}_{\eta}-2m^{2}_{\Sigma^*}m^{2}_{\eta})^{1/2}\,.
\end{eqnarray}
To avoid the over-calculation of the off-shell contribution
from the exchanged meson~\cite{Li:1996yn},
the form factor $F(q_{2}^2)$ has been introduced in Eq.~(\ref{Aab2}),
together with
$P_{\mu\nu}$ that sums over 
the 3/2-spin for $\Sigma^{*+}$, given by~\cite{Cheng:2004ru,Toki:2007ab}
\begin{eqnarray}\label{Aab3}
&&F(q_{2}^2)=\frac{c_\Lambda^{2}-m^{2}_{\pi}}{c_\Lambda^{2}-q^{2}_{2}\;}\,,\nonumber\\
&&P_{\mu\nu}
=(\slashed{q_{1}}+m_{\Sigma^*})(-g_{\mu\nu}
+\frac{1}{3}\gamma_{\mu}\gamma_{\nu}
-\frac{q_{1\mu}\gamma_{\nu}-q_{1\nu}\gamma_{\mu}}{3m_{\Sigma^*}}
+\frac{2q_{1\mu}q_{1\nu}}{3 m_{\Sigma^*}^{2}})\,,
\end{eqnarray}
where $c_\Lambda$ is the cut-off parameter.

\newpage
To reduce Eq.~(\ref{Aab2}), we need the following identities:
\begin{eqnarray}\label{Aab4}
\bar{u}_\Lambda\slashed{q_{1}}\slashed{q_{2}} u_{\Lambda_c}&=&
\bar{u}_\Lambda(-2q_{1}\cdot p_{2}+m_{\Sigma^*}^{2}+\slashed{q_{1}}m_{\Lambda}) u_{\Lambda_c}\,,
\nonumber\\
\bar{u}_\Lambda\slashed{q_{1}}\slashed{q_{3}} u_{\Lambda_c}&=&
\bar{u}_\Lambda(-m_{\Sigma^*}^{2}+\slashed{q_{1}}m_{\Lambda_{c}}) u_{\Lambda_c}\,,
\nonumber\\
\bar{u}_\Lambda\slashed{q_{1}}\slashed{q_{2}}\slashed{q_{3}} u_{\Lambda_c}&=&
\bar{u}_\Lambda[(m_{\Sigma^*}^{2}-2q_{1}\cdot p_{2})m_{\Lambda_{c}}-m_{\Lambda}m_{\Sigma^*}^{2}
\nonumber\\
&+&
\slashed{q_{1}}(m_{\Lambda_{c}}m_{\Lambda}-m_{\Sigma^*}^{2}+2q_{1}\cdot p_{2})] u_{\Lambda_c}\,,
\nonumber\\
 \bar{u}_\Lambda\gamma_{5}\slashed{q_{1}}\slashed{q_{2}} u_{\Lambda_c}&=&\bar{u}_\Lambda\gamma_{5}(-2q_{1}\cdot p_{2}+m_{\Sigma^*}^{2}-\slashed{q_{1}}m_{\Lambda}) u_{\Lambda_c}\,,\nonumber\\
 \bar{u}_\Lambda\gamma_{5}\slashed{q_{1}}\slashed{q_{3}} u_{\Lambda_c}&=&\bar{u}_\Lambda\gamma_{5}(-m_{\Sigma^*}^{2}+\slashed{q_{1}}m_{\Lambda_{c}}) u_{\Lambda_c}\,,\nonumber\\
 \bar{u}_\Lambda\gamma_{5}\slashed{q_{1}}\slashed{q_{2}}\slashed{q_{3}} u_{\Lambda_c}&=&\bar{u}_\Lambda\gamma_{5}[(m_{\Sigma^*}^{2}-2q_{1}\cdot p_{2})m_{\Lambda_{c}}+m_{\Lambda}m_{\Sigma^*}^{2}\nonumber\\
 &+&\slashed{q_{1}}(-m_{\Lambda_{c}}m_{\Lambda}-m_{\Sigma^*}^{2}+2q_{1}\cdot p_{2})] u_{\Lambda_c}\,.
\end{eqnarray}
In Eq.~(\ref{Aab2}), the integration results in
\begin{eqnarray}\label{Aab5}
\int d\Omega  \bar{u}_\Lambda\slashed{q_{1}} u_{\Lambda_c}&=&
\int d\Omega\bigg[E_{1}-\frac{|\vec{q}_{1}|}{|\vec{p}_{2}|}(E_{\Lambda}-m_{\Lambda})\cos\theta\bigg]
\bar{u}_\Lambda u_{\Lambda_c}\,,
\nonumber\\
\int d\Omega  \bar{u}_\Lambda\gamma_{5}\slashed{q_{1}} u_{\Lambda_c}&=&
\int d\Omega \bigg[E_{1}-\frac{|\vec{q}_{1}|}{|\vec{p}_{2}|}(E_{\Lambda}+m_{\Lambda})\cos\theta\bigg]
\bar{u}_\Lambda\gamma_{5} u_{\Lambda_c}\,.
\end{eqnarray}
We hence obtain
\begin{eqnarray}\label{amp2}
{\cal M}(\Lambda_{c}^+\to \Lambda a_{0}^{+})=\bar{u}_\Lambda(A-B\gamma_{5}) u_{\Lambda_c}\,,
\end{eqnarray}
where $A$ and $B$ are given by
\begin{eqnarray}
A&=&\frac{|\vec{q}_{1}|}{16\pi}
\frac{g_{\Lambda_{c}\Sigma^{*}\eta}g_{\Sigma^{*}\Lambda\pi}g_{a_{0}\eta\pi}}{m_{\Lambda_c} m_\eta m_\pi}
\int d \cos\theta
\frac{\alpha_{0}+\alpha_{1}\cos\theta+\alpha_{2}\cos^{2}\theta}
{(q^{2}_{2}-m^{2}_{\pi})(q^{2}_{2}-c_\Lambda^{2})^{2}}\,,
\nonumber\\
B&=&\frac{|\vec{q}_{1}|}{16\pi}
\frac{g_{\Lambda_{c}\Sigma^{*}\eta}g_{\Sigma^{*}\Lambda\pi}g_{a_{0}\eta\pi}}{m_{\Lambda_c} m_\eta m_\pi}
\int d \cos\theta
\frac{\beta_{0}+\beta_{1}\cos\theta+\beta_{2}\cos^{2}\theta}
{(q^{2}_{2}-m^{2}_{\pi})(q^{2}_{2}-c_\Lambda^{2})^{2}}\,,
\end{eqnarray}
with
\begin{eqnarray}
\alpha_{0}&=&
\frac{1}{12 m_{\Sigma^*}^{2} m_{\Lambda_{c}}^{2}}
\{E_{\Lambda}[-m_{\Sigma^*}^{6}-7m_{\Sigma^*}^{5}m_{\Lambda_{c}}+m_{\Sigma^*}^{4}(m^{2}_{\Lambda_{c}}+3m_{\eta}^{2})
+6m_{\Sigma^*}^{3}(m_{\eta}^{2}m_{\Lambda_{c}}+m_{\Lambda_{c}}^{3})\nonumber\\
&+&m_{\Sigma^*}^{2}(m_{\Lambda_{c}}^{4}+2m_{\Lambda_{c}}^{2}m_{\eta}^{2}-3m_{\eta}^{4})+(m_{\Lambda_{c}}^{2}-m_{\eta}^{2})^{3}]\nonumber\\
&+&m_{\Sigma^*}m_{\Lambda_{c}}m_{\Lambda}[-3m_{\Sigma^*}^{4}+2m_{\Sigma^*}^{2}(m^{2}_{\Lambda_{c}}+m_{\eta}^{2})+(m_{\Lambda_{c}}^{2}-m_{\eta}^{2})^{2}]\}\,,\nonumber\\
\alpha_{1} &=& \frac{|\vec{q}_{1}|}{6m_{\Sigma^*}^{2}m_{\Lambda_{c}}|\vec{p}_{2}|}\{2E^{2}_{\Lambda}[m_{\Sigma^*}^{4}+m_{\Sigma^*}^{3}m_{\Lambda_{c}}-2m_{\Sigma^*}^{2}m^{2}_{\Lambda_{c}}
+m_{\Sigma^*}(m_{\eta}^{2}m_{\Lambda_{c}}-m_{\Lambda_{c}}^{3})+(m_{\Lambda_{c}}^{2}-m_{\eta}^{2})^{2}]\nonumber\\
&-&E_{\Lambda}m_{\Lambda}[m_{\Sigma^*}^{4}-2m_{\Sigma^*}^{2}(m^{2}_{\Lambda_{c}}+m_{\eta}^{2})+(m_{\Lambda_{c}}^{2}-m_{\eta}^{2})^{2}]\nonumber\\
&-&m^{2}_{\Lambda}[m_{\Sigma^*}^{4}+6m_{\Sigma^*}^{3}m_{\Lambda_{c}}+2m_{\Sigma^*}^{2}(m^{2}_{\Lambda_{c}}-m_{\eta}^{2})+2m_{\Sigma^*}(m_{\eta}^{2}m_{\Lambda_{c}}-m_{\Lambda_{c}}^{3})+(m_{\Lambda_{c}}^{2}-m_{\eta}^{2})^{2}]\}\,,\nonumber\\
\alpha_{2} &=&
-|\vec{q}_{1}|^{2}(E_{\Lambda}-m_\Lambda)\frac{m_{\Sigma^*}^{2}-m_{\Sigma^*}m_{\Lambda}+m_{\Lambda}^{2}-m_{\eta}^{2}}{3m_{\Sigma^*}^{2}}\,,
\end{eqnarray}
and
\begin{eqnarray}
\beta_{0}&=&
\frac{m_{\Sigma^*}^{4}-2m_{\Sigma^*}^{2}(m^{2}_{\Lambda_{c}}
+m_{\eta}^{2})+(m_{\Lambda_{c}}^{2}-m_{\eta}^{2})^{2}}{12 m_{\Sigma^*}^{2} m_{\Lambda_{c}}^{2}}
\nonumber\\&\times&
[E_{\Lambda}(m_{\Sigma^*}^{2}-3m_{\Sigma^*}m_{\Lambda_{c}}+m^{2}_{\Lambda_{c}}-m_{\eta}^{2})
-m_{\Sigma^*}m_{\Lambda_{c}}m_{\Lambda}]\,,\nonumber\\
\beta_{1} &=&
-\frac{|\vec{q}_{1}|}{6m_{\Sigma^*}^{2}m_{\Lambda_{c}}|\vec{p}_{2}|}(m_{\Sigma^*}^{2}-2m_{\Sigma^*}m_{\Lambda}+m_{\Lambda}^{2}-m_{\eta}^{2})[2E^{2}_{P_{2}}(m_{\Sigma^*}^{2}+m_{\Sigma^*}m_{\Lambda_{c}}+m^{2}_{\Lambda_{c}}-m_{\eta}^{2})\nonumber\\
&+&E_{\Lambda}m_{\Lambda}(m_{\Sigma^*}^{2}+2m_{\Sigma^*}m_{\Lambda_{c}}+m^{2}_{\Lambda_{c}}-m_{\eta}^{2})-m^{2}_{\Lambda}(m_{\Sigma^*}^{2}+m^{2}_{\Lambda_{c}}-m_{\eta}^{2})]\,,\nonumber\\
\beta_{2} &=&
|\vec{q}_{1}|^{2}(E_{\Lambda}+m_\Lambda)\frac{m_{\Sigma^*}^{2}+m_{\Sigma^*}m_{\Lambda}+m_{\Lambda}^{2}-m_{\eta}^{2}}{3m_{\Sigma^*}^{2}}\,.
\end{eqnarray}
Using Eq.~(\ref{amp2}), we present the decay width:
\begin{eqnarray}
\Gamma(\Lambda^{+}_{c}\to \Lambda a_{0}^{+})&=&
\frac{|\vec{p}_{2}|}{8\pi m_{\Lambda_c}^2}\bigg[
(m_+^2 -m_{a_0}^2)|A|^2+(m_-^2 -m_{a_0}^2)|B|^2\bigg]\,,
\end{eqnarray}
to be used in the numerical analysis,
where $m_{\pm}=m_{\Lambda_{c}}\pm m_{\Lambda}$.

\section{Numerical Results and Discussions}
To perform the numerical analysis,
we determine $g_{\Sigma^{*}\Lambda\pi}$ and $g_{a_0\eta\pi}$ from
${\cal B}(\Sigma^{*+}\to \Lambda\pi^{+}) = (87.0\pm1.5)\%$
and  ${\cal B}(a_0\to \eta\pi) = (84.5\pm 1.7)\%$, respectively~\cite{pdg,Cheng:2013fba}.
By following the topological scheme with the $SU(3)$ flavor symmetry~\cite{Hsiao:2020iwc},
we extract $a_E$. Hence,
we present the coupling constants as
\begin{eqnarray}\label{ggg}
&&
g_{\Sigma^{*}\Lambda\pi}=1.30\pm0.05\,,\;
g_{a_0\eta\pi}=(2.53\pm0.03)~\text{GeV}\,,\;
\nonumber\\
&&
a_E=(0.065\pm 0.014)~\text{GeV}^2\,.
\end{eqnarray}
The cutoff parameter $c_\Lambda$ is not well controlled.
The estimation gives
$c_\Lambda=\lambda \Lambda_{\text{QCD}}+m_\pi$,
where $\Lambda_{\text{QCD}}=0.22$~GeV,
and $\lambda$ is a free parameter in the range of 1 to 3~\cite{Cheng:2004ru,Ke:2020uks,Ke:2010aw}.
On the other hand,
$c_\Lambda$ can be determined by the data. For example,
one has used $c_\Lambda=(0.8,1.0)$~GeV to demonstrate
the rescattering contribution to $\Lambda_c^+\to \Sigma^0 \pi^+$,
in order to interpret the branching fraction and up-down asymmetry ($\alpha$)~\cite{Ke:2020uks}.
Since $\Lambda_c^+\to \Sigma^+ \pi^0,\Sigma^0\pi^+$
respect the isospin symmetry~\cite{Geng:2017esc},
we average the experimental values as
${\cal B}(\Lambda_c^+\to \Sigma^{0(+)}\pi^{+(0)})=(1.3\pm 0.1)\%$ and
${\cal \alpha}(\Lambda_c^+\to \Sigma^{0(+)}\pi^{+(0)})=-0.6\pm 0.2$~\cite{pdg},
which correspond to the more restricted cut-off parameter $c_\Lambda=0.95\pm 0.05$.

\newpage
Consequently, we obtain
\begin{eqnarray}\label{Aab7}
&&{\cal B}(\Lambda^{+}_{c}\to \Lambda a_{0}^{+})
=(1.7^{+2.8}_{-1.0}\pm 0.3)\times 10^{-3}\,,
\end{eqnarray}
where the first error comes from $\delta c_\Lambda$,
and the second one combines the uncertainties of
the strong coupling constants and $a_E$ in Eq.~(\ref{ggg}).
Since ${\cal B}(\Lambda^{+}_{c}\to \Lambda a_{0}^{+})$
is predicted ten times larger than that in~\cite{Sharma:2009zze},
it demonstrates that the final state interaction dominates the contribution,
instead of the factorizable and non-factorizable effects.

According to ${\cal B}(\Lambda_c^+\to \Lambda^*\pi^+)\simeq
{\cal B}(\Lambda_c^+\to \Sigma^{*+}\eta)$~\cite{Lee:2020xoz,pdg},
the rescattering decay $\Lambda_c^+\to \Lambda^*\pi^+\to\Lambda a_0^+$
with the $\eta$ exchange can be another sizeable contribution.
As the test,
${\cal B}(\Lambda_c^+\to \Lambda^*\pi^+\to\Lambda a_0^+)=
(5.3^{+1.6}_{-1.2})\times10^{-5}$ is calculated to be much smaller than
${\cal B}(\Lambda_c^+\to \Sigma^*\eta\to\Lambda a_0^+)$, due to that
$g_{\Lambda^*\Lambda\eta}$ extracted from ${\cal B}(\Lambda^*\to\Lambda\eta)$
is much smaller than $g_{\Sigma^{*}\Lambda\pi}$~\cite{Lee:2020xoz,pdg}.
Besides, $m_\eta\gg m_\pi$ causes a serious suppression in the integration, where
$\eta$ behaves as the exchanged particle between $\Lambda^*$ and $\pi^+$
in the triangle loop.

The triangle singularity (TS) is commonly used
to enhance the triangle rescattering decay~\cite{Liu:2019dqc}.
In our case,
$\Lambda_{c}^+\to\Sigma^*\eta\to\Lambda a_{0}^{+}$
with $m_{\Sigma^*}\simeq 1.4$~GeV is not located in the so-called physical region,
such that the TS cannot be induced~\cite{TS,Guo:2019twa}. By contrast, 
$\Lambda_{c}^+\to\Sigma'\eta\to\Lambda a_{0}^{+}$ can fulfil the TS condition, 
where $\Sigma'$ as a heavier higher-wave $\Sigma$ baryon 
can be $\Sigma(1620)$, $\Sigma(1660)$ or $\Sigma(1670)$.
As a result, the triangle loop with $\Sigma'$  
leads to the integration three times larger than that with $\Sigma^*$. Similarly, 
$\Lambda_c^+\to N^{\prime}\bar K^0\to \Lambda a_0^+$ with $K^+$ exchange
can also be enhanced by the TS condition,
where $N^{\prime}$ can be $N(1700)$, $N(1710)$ or $N(1720)$.

Despite of the TS enhancement,
it is not necessary that
${\cal B}(\Lambda_c^+\to \Sigma'\eta(N^{\prime}\bar K^0)\to\Lambda a_0^+)$
can be as large as ${\cal B}(\Lambda_c^+\to \Sigma^*\eta\to\Lambda a_0^+)$.
The estimation, however, seems difficult without the extractable weak coupling constants 
from ${\cal B}(\Lambda_c^+\to \Sigma'\eta,N^{\prime}\bar K^0)$ that
have not been observed yet. For a different estimation, 
we derive 
\begin{eqnarray}\label{ext1}
&&
{\cal B}(\Lambda_c^+\to \Sigma'\eta,\Sigma'\to\Lambda\pi^+)
\simeq8{\cal B}(\Lambda_c^+\to \Sigma' \eta\to\Lambda a_0^+)\,,\nonumber\\
&&
{\cal B}(\Lambda_c^+\to N'\bar K^0,N'\to\Lambda K^+)
\simeq5{\cal B}(\Lambda_c^+\to N'\bar K^0\to\Lambda a_0^+)\,,
\end{eqnarray}
which is by using the fact that 
the resonant and rescattering decays can be related with the same coupling constants.
In addition, the two inequalities can be useful,
given by
\newpage
\begin{eqnarray}\label{ext2}
&&
{\cal B}(\Lambda_c^+\to (N'\bar K^0,\Sigma^{\prime} \eta,\Sigma^* \eta)\to\Lambda a_0^+)
{\cal B}(a_0^+\to\eta\pi^+)\nonumber\\
&&
+{\cal B}(\Lambda_c^+\to \Sigma^{\prime}\eta,\Sigma^{\prime}\to\Lambda\pi^+)
\le {\cal B}'(\Lambda_c^+\to \Lambda\eta\pi^+)\,,\nonumber\\
&&
{\cal B}(\Lambda_c^+\to N'\bar K^0,N'\to\Lambda K^+)\le
{\cal B}(\Lambda_c^+\to \Lambda K^+\bar K^0)\,,
\end{eqnarray}
where ${\cal B}'(\Lambda_c^+\to \Lambda\eta\pi^+)=(4.4\pm 1.8)\times 10^{-3}$ 
has excluded the resonant contributions in Eq.~(\ref{data2}) from 
${\cal B}(\Lambda_c^+\to \Lambda\eta\pi^+)$ in Eq.~(\ref{data1}), and
${\cal B}(\Lambda_c^+\to\Lambda K^+\bar K^0)=(4.1\pm 1.2)\times 10^{-3}$~\cite{pdg}.
With the two inequalities combined as one, where
${\cal B}(\Lambda_c^+\to N'\bar K^0,N'\to\Lambda K^+)$ and 
${\cal B}(\Lambda_c^+\to \Sigma'\eta,\Sigma'\to\Lambda\pi^+)$
are replaced by those in Eq.~(\ref{ext1}), we obtain
\begin{eqnarray}
&&
{\cal B}(\Lambda_c^+\to \Sigma(1620,1660,1670) \eta\to\Lambda a_0^+)\nonumber\\
&&+
{\cal B}(\Lambda_c^+\to N(1700,1710,1720)\bar K^0\to\Lambda a_0^+)
\lesssim 1.0\times 10^{-3}\,.
\end{eqnarray}
Note that the contribution from 
${\cal B}(\Lambda_c^+\to \Sigma^* \eta\to\Lambda a_0^+)$ has been removed
with the value in Eq.~(\ref{Aab7}).
Therefore, individually ${\cal B}(\Lambda_c^+\to \Sigma'\eta\to\Lambda a_0^+)$ and 
${\cal B}(\Lambda_c^+\to N'\bar K^0\to\Lambda a_0^+)$ can be a few$\times 10^{-4}$ at most.
On the other hand, ${\cal B}(\Lambda^{+}_{c}\to \Lambda a_{0}^{+})\simeq 2\times 10^{-3}$
is as large as the measured ${\cal B}(\Lambda_c^+\to p f_0)=(3.5\pm 2.3)\times 10^{-3}$,
promising to be observed in the near future.

\section{Conclusions}
In the summary, we have studied the $\Lambda^{+}_{c}\to\Lambda a_{0}^{+}$ decay,
inspired by the recent observation of $\Lambda_c^+\to \Lambda\eta\pi^+$,
which has hinted a possible signal for
the resonant $\Lambda_c^+\to \Lambda a_0^+, a_0^+\to \eta\pi^+$ process.
The calculation based on the factorization and pole model
has given its branching fraction as small as $1.9\times 10^{-4}$,
presenting the limited factorizable and non-factorizable effects.
On the other hand, we have found that the final state interaction
can significantly contribute to $\Lambda_c^+\to \Lambda a_0^+$,
where the $\Lambda_c^+\to \Sigma^{*+}\eta$ decay is followed by
the $\Sigma^{*+}$ and $\eta$ rescattering.
With the exchange of a charged pion,
$\Sigma^{*+}$ and $\eta$ are transformed as $\Lambda$ and $a_0^+$, respectively.
Consequently, we have predicted
${\cal B}(\Lambda^{+}_{c}\to\Lambda a_{0}^{+})=(1.7^{+2.8}_{-1.0}\pm 0.3)\times 10^{-3}$
an order of magnitude larger than the previous calculation,
which is promising to be observed at the BESIII, BELLEII and LHCb experiments.

\newpage
\section*{ACKNOWLEDGMENTS}
YKH was supported in part by NSFC (Grant No.~11675030).
YY was supported in part by NSFC (Grant No.~11905023) and
CQCSTC (cstc2020jcyj-msxmX0555, cstc2020jcyj-msxmX0810).


\begin{thebibliography}{99}
\bibitem{review}
Please consult Ref.~\cite{pdg} for a review.

\bibitem{pdg}
P.A.~Zyla \textit{et al.} [Particle Data Group],
PTEP \textbf{2020}, 083C01 (2020).

\bibitem{Barlag:1990yv}
S.~Barlag \textit{et al.} [ACCMOR],
Z. Phys. C \textbf{48}, 29 (1990).

\bibitem{Wang:2020pem}
Z.~Wang, Y.Y.~Wang, E.~Wang, D.M.~Li and J.J.~Xie,
Eur.\ Phys.\ J.\ C {\bf 80}, 842 (2020). 

\bibitem{Li:2020fqp}
H.S.~Li, L.L.~Wei, M.Y.~Duan, E.~Wang and D.M.~Li,
arXiv:2009.08600 [hep-ph].

\bibitem{Lee:2020xoz}
J.Y.~Lee {\it et al.} [Belle Collaboration],
Phys.\ Rev.\ D {\bf 103}, 052005 (2021). 

\bibitem{Xie:2016evi}
J.J.~Xie and L.S.~Geng,
Eur.\ Phys.\ J.\ C {\bf 76}, 496 (2016). 

\bibitem{Sharma:2009zze}
A.~Sharma and R.C.~Verma,
J.\ Phys.\ G {\bf 36}, 075005 (2009).

\bibitem{Zhao:2018mov}
H.J.~Zhao, Y.L.~Wang, Y.K.~Hsiao and Y.~Yu,
JHEP {\bf 2002}, 165 (2020). 

\bibitem{Kohara:1991ug}
Y.~Kohara, 
Phys.\ Rev.\ D {\bf 44}, 2799 (1991).

\bibitem{Hsiao:2020iwc}
Y.K.~Hsiao, Q.~Yi, S.T.~Cai and H.J.~Zhao,
Eur.\ Phys.\ J.\ C {\bf 80}, 1067 (2020). 

\bibitem{Pan:2020qqo}
J.~Pan, Y.K.~Hsiao, J.~Sun and X.G.~He,
Phys.\ Rev.\ D {\bf 102}, 056005 (2020).

\bibitem{Liu:2019dqc}
X.H.~Liu, G.~Li, J.J.~Xie and Q.~Zhao,
Phys.\ Rev.\ D {\bf 100}, 054006 (2019). 

\bibitem{TS} R.J.~Eden, P.V.~Landsho, D.I.~Olive, and J.C.~Polkinghorne,
The Analytic S-Matrix (Cambridge University Press, Cambridge, 1966).

\bibitem{Guo:2019twa}
F.K.~Guo, X.H.~Liu and S.~Sakai,
Prog.\ Part.\ Nucl.\ Phys.\  {\bf 112}, 103757 (2020). 

\bibitem{Hsiao:2019ait}
Y.K.~Hsiao, Y.~Yu and B.C.~Ke,
Eur. Phys. J. C \textbf{80}, 895 (2020). 

\bibitem{Ke:2020uks}
H.W.~Ke and X.Q.~Li,
Phys.\ Rev.\ D {\bf 102}, 113013 (2020). 

\bibitem{Hsiao:2020gtc}
Y.K.~Hsiao, L.~Yang, C.C.~Lih and S.Y.~Tsai,
Eur.\ Phys.\ J.\ C {\bf 80}, 1066 (2020). 

\bibitem{Xie:2017xwx}
J.J.~Xie and L.S.~Geng,
Phys.\ Rev.\ D {\bf 95}, 074024 (2017). 

\bibitem{Gutsche:2018utw}
T.~Gutsche, M.A.~Ivanov, J.G.~Korner and V.E.~Lyubovitskij,
Phys. Rev. D {\bf 98}, 074011 (2018). 

\bibitem{FKS}
T.~Feldmann, P.~Kroll and B.~Stech,
Phys.\ Rev.\ D {\bf 58}, 114006 (1998); Phys.\ Lett.\ B {\bf 449}, 339 (1999).

\bibitem{Li:1996yn}
X.Q.~Li, D.V.~Bugg and B.S.~Zou,
Phys.\ Rev.\ D {\bf 55}, 1421 (1997).

\bibitem{Cheng:2004ru}
H.Y.~Cheng, C.K.~Chua and A.~Soni,
Phys. Rev. D \textbf{71},014030 (2005). 

\bibitem{Toki:2007ab}
H.~Toki, C.~Garcia-Recio and J.~Nieves,
Phys. Rev. D \textbf{77}, 034001 (2008). 

\bibitem{Cheng:2013fba}
H.Y.~Cheng, C.K.~Chua, K.C.~Yang and Z.Q.~Zhang,
Phys. Rev. D \textbf{87}, 114001 (2013). 

\bibitem{Ke:2010aw}
H.W.~Ke, X.Q.~Li and X.~Liu,
Phys.\ Rev.\ D {\bf 82}, 054030 (2010). 

\bibitem{Geng:2017esc}
C.Q.~Geng, Y.K.~Hsiao, Y.H.~Lin and L.L.~Liu,
Phys.\ Lett.\ B {\bf 776}, 265 (2018). 

\end{thebibliography}
\end{document}